# Friction Tubes to Generate Nanobubble Ozone Water with an Increased Half-Life for Virucidal Activity

Suk-Joo Byun,[1] A-Ram You,[2] Tae Seok Park,[3] Chang-Hee Park,[1] Dae-Hyun Choi,[1] Eun-Hee Jun,[1] Young-Ho Yoo,[1] Taekeun Yoo[1]

1 Fawoo Nanotech Corporation, Bucheon-si, Gyeonggi-do, Republic of Korea
2 School of Architecture, Kumoh National Institute of Technology, Gumi, Gyeongbuk, Republic of Korea
3 Dmax Co., Ltd., 89 Hosan-ro, Dalseo-gu, Daegu, Republic of Korea
Correspondence: Young-Ho Yoo, fawoonano@naver.com; Taekeun Yoo, fawoo2@snu.ac.kr

## Abstract

Nanobubbles and related technologies are expected to be highly utilized in water resource-based industries such as water purification, crops, horticulture, medicine, bio, and sterilization. Ozone, a chemical with high sterilizing power, is known as a natural substance that is reduced to oxygen and water after reacting with pollutants. Ozone water, which is generated by dissolving ozone in water, has been used in various industrial sectors such as medical care, food, and environment. Due to the unstable molecular state of ozone, however, it is difficult to produce, use, and supply ozone at industrial sites in a stable manner. This study proposed a method for constructing a system that can generate high-concentration ozone water in large quantities using low power in real time and maintaining the concentration of the generated ozone water over the long term. Friction tubes (called "nanotube") played a key role to generate nanobubble ozone water with an increased half-life for virus killing activity. In addition, the safety of ozone water during its spray into the air was explained, and virucidal activity test cases for the influenza A (H1N1/A/PR8) and COVID-19 (SARS-CoV-2) virus using high-concentration ozone water as well as its technical efficacy were described.

## I. Introduction

### 1. Research Background

Nanobubbles and related technologies are expected to be highly utilized in water resource-based industries such as water purification, crops, horticulture, medicine, bio, and sterilization.[1] A recent study introduced a friction tube to generate massive volume of nanobubble water.[2] In order for these technologies to be used in actual fields, additional application technologies must be combined.

The ozone (Ozone-$O^3$) molecules concentrated in the ozone layer within the earth's stratosphere protect life on earth by blocking the intense ultraviolet rays from the sun. Such gaseous ozone also exists in the air of high mountains, coasts, and forests. It is also a source of feeling refreshed. Ozone water, which contains ozone with high oxidizing power, has also been used for various purposes such as microbial removal, sterilization, and cleaning.[3]

Ozone is a natural substance that is reduced to oxygen and water after reacting with pollutants. The presence of a large amount of gaseous ozone in the air, however, provides discomfort and has harmful effects on the human body, such as bronchial stimuli and lung function degradation. Gaseous ozone is also difficult to use due to its instability and safety problem when leaked.

On the other hand, ozone water, which is generated by dissolving ozone in water, can fully exhibit the sterilizing power of ozone due to the adhesion of water.[4] While ozone water is generated by mixing ozone with water, ozone is not dissolved and released immediately as ozone gas if added into water in large bubbles.[5] This makes it impossible to maintain a constant ozone concentration in ozone water and obtain desired effects, such as purifying and sterilizing power.[6] Since the ozone gas separated from water without being collected has unique odor and toxicity, a separate ozone gas treatment system must be prepared. A previous study showed that ozone could show the effect of ozone treatment on SARS-CoV-2.[7]

To produce ozone water, technologies, such as turbine contactor, porous diffuser, and injector methods, are generally used. Turbine agitation rotates water using a turbine to mix it with ozone when the air containing ozone is injected at the bottom of the water tank through a pipe.[8] For the porous diffuser method, the air containing ozone is injected at the bottom of the water tank through a pipe and the ozone is refined through the small pores of a disk plate. In the case of the injector method,[9] the gas containing ozone is absorbed into the injector in the vacuum state by the pump pressure and air-blown with water. A method using an expensive electrocatalyst has also been introduced.[10]

These conventional technologies require a large space for installation, and the generated ozone water has a low ozone concentration and a short half-life of 20 to 25 minutes. Due to the short half-life and low-concentration problem, ozone must be produced and used immediately at the site.

When ozone water is generated using tap water, the ozone concentration becomes low despite the input of high-concentration ozone due to the reaction with the inorganic matter included in the water and decomposition. Even after the generation of ozone water, the ozone concentration half-life decreases because ozone continuously reacts with the organic matter contained in the water and decomposes. In other words, the half-life or extinction rate of ozone is closely affected by the presence or amounts of the organic and inorganic matter contained in the water.

It is possible to increase the concentration and half-life of ozone by generating ozone water by dissolving ozone in distilled water as an alternative. This method, however, is used only in some limited fields, including medical care, because the use of distilled water incurs considerable additional cost.

## 2. Research Purpose

To address these problems, ozone was dissolved in the nanobubbles generated using friction tubes in this study to produce high-concentration ozone water in large quantities using low power in real time. Tap water and distilled water were used for the generation of high-concentration ozone water. Each ozone concentration was examined, and the method of maintaining the concentration over the long term was tested according to the temperature. If the ozone concentration half-life is significantly increased, the applicability of ozone will be further improved.

When high-concentration ozone water was sprayed into the air for sterilization, it was examined whether the stability of atmospheric environmental standards was met. Influenza A and COVID-19 virucidal activity testing was conducted using high-concentration ozone water to examine its applicability as a disinfectant.

It is expected that the nanobubble ozone water generated using the proposed method will have no risk of chemical secondary pollution and significantly extend the utilization range of ozone water while achieving the treatment efficiency increase, processing time reduction, and cost saving compared to conventional cleaning or disinfection methods.

## II. Method to Generate Nanobubble Ozone Water and Increase Half-Life

### 1. Nanobubble ozone water production

#### 1.1 Experimental setup

For the generation of nanobubbles, which is technologically important in this study, the "nanotubes" were used (detailed information is described in a previous study[2]). In the study, the concepts of (1) the scale factor (= friction integration) and (2) the effective friction constant are described.

$$\text{Scale factor (=friction integration)} = \frac{\text{Inner circumference of the cross section(mm)}}{\text{Cross-sectional area of the flow path (mm}^2)} \quad \text{---------------------- (1)}$$

Effective friction constant = scale factor * nanotube length (m) ---------------------------------- (2)

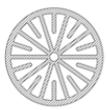

Table. 2-1 Scale factor according to the nanotube cross section

#### 1.2 Sample preparation

In the "standards and specifications of food additives - notice no. 2023-60 (September 20, 2023) announced by the Ministry of Food and Drug Safety, ozone water is defined as an aqueous solution with ozone as its main component, which is obtained by dissolving the ozone gas produced from an ozone generator. It is also specified that the ozone content must be 1.0 mg/L (1.0 ppm) or higher.

In this study, ozone water samples were generated using tap water and distilled water. For tap water, purification methods and chemical components have been standardized worldwide. Tap water was used after installing a filtration system because it has uniform quality after disinfection and sterilization and is easy to use.

In the case of distilled water, type 1 (ultra-pure water) was used among the four standards (D1193-06 (2018), Standard Specification for Reagent Water) determined by the American Society for Testing and Materials (ASTM).

#### 1.3 Ozone concentration measurement system

An ozone portable colorimeter (model: Q-03-2, SINSCHE Technology Co., Ltd) was used to measure the ozone concentration in the generated ozone water. As for the measurement method, the generated ozone water sample is placed in a dedicated glass bottle (12.5 ml) and well mixed after adding a reagent for measurement (R1). The sample bottle is then placed into the measurement system to read the displayed concentration value. To improve the reliability of the analysis results, measurements were performed without changing the setting of the system and dilution.

### 2. Low-concentration nanobubble ozone water production and ozone concentration change - tap water

#### 2.1 Experimental setup

The system to generate nanobubble ozone water mainly consists of a sample tank, an ozone generator, a pump, and a nanotube hose. Its specifications and detailed configuration are described below.

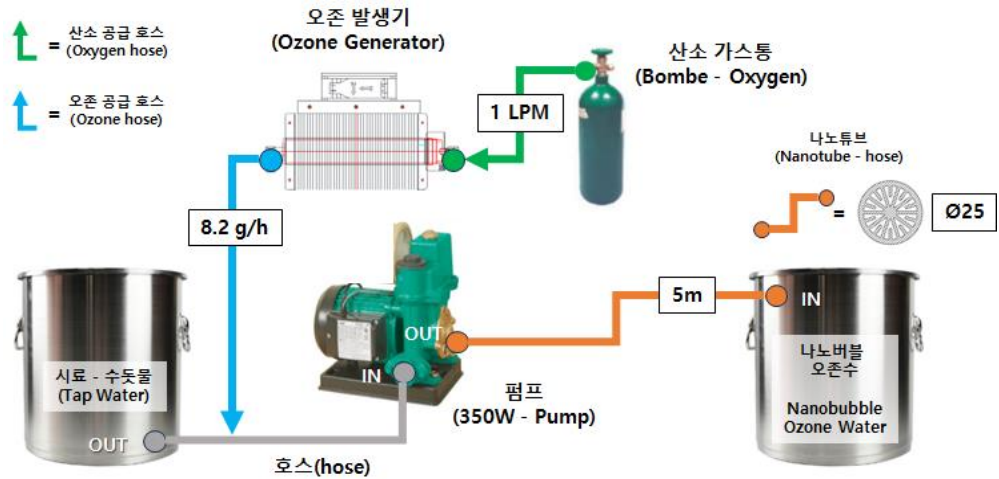

Fig. 2-1 Nanobubble ozone water generation system – for general ozone water production

To produce general ozone water, the sample tank made of SUS 304 (stainless steel, Ni 8~11%, Cr 18~20%) was filled with water and the water pump with a rated output of 350 W, a pumping capacity of 26 ℓ/min, and a rated power of 220V/60Hz (model: PW-S354M, Wilo Pumps Ltd.) was prepared to increase flow velocity as shown in Fig. 2-1. The input of the pump was connected to the output of the sample tank using a silicon hose. A nanotube hose with a scale factor of 1.4, an effective friction constant of 7.0, an outer diameter of Ø25, and a length of 5 m was connected to the output of the pump. The opposite end of the nanotube hose was connected to another sample tank made of SUS 304. After confirming a shear pressure of 1.8 bars from the nanotube hose, nanobubble ozone water was produced while the ozone generated through the ozone generator was injected at a rate of 8.2 g/h.

The ozone generator used was produced by Ozone Tech (model: OG-10R). To increase the concentration of the ozone generated from the ozone generator, oxygen was supplied into the ozone generator at a rate of 1.0 LPM using the oxygen bombe.

## 2.2 Production of general ozone water with a concentration of 1 ppm or higher

The concentration of the ozone water generated at a tap water temperature of approximately 20°C was measured over time. For the reliability of the measurement results, the same sample was placed in the seven 12.5ml glass bottles prepared, and measurements were performed through reactions with the reagent (R1) in new bottles every hour at room temperature. 1.3 ppm was confirmed at the beginning, and it was reduced to a half in approximately 50 minutes. After five hours, no more measurement was performed because insignificant values were obtained.

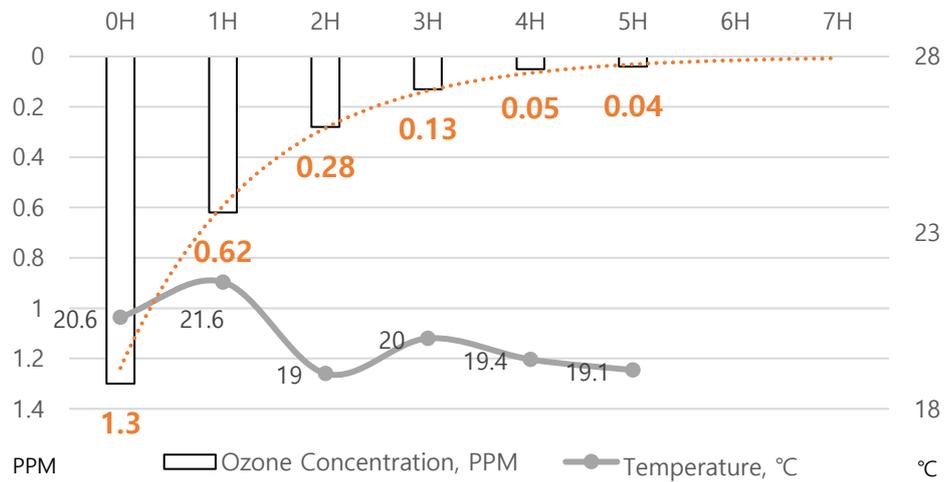

Fig. 2-2 Ozone water concentration change – tap water, room temperature 20°C / room temperature

## 3. High-concentration nanobubble ozone water production and ozone concentration change - tap water

### 3.1 Experimental setup

High-concentration nanobubble ozone water was generated using tap water and then half-life was examined and compared. While the laboratory temperature was maintained at 24°C, a system was constructed for the experiment as follows.

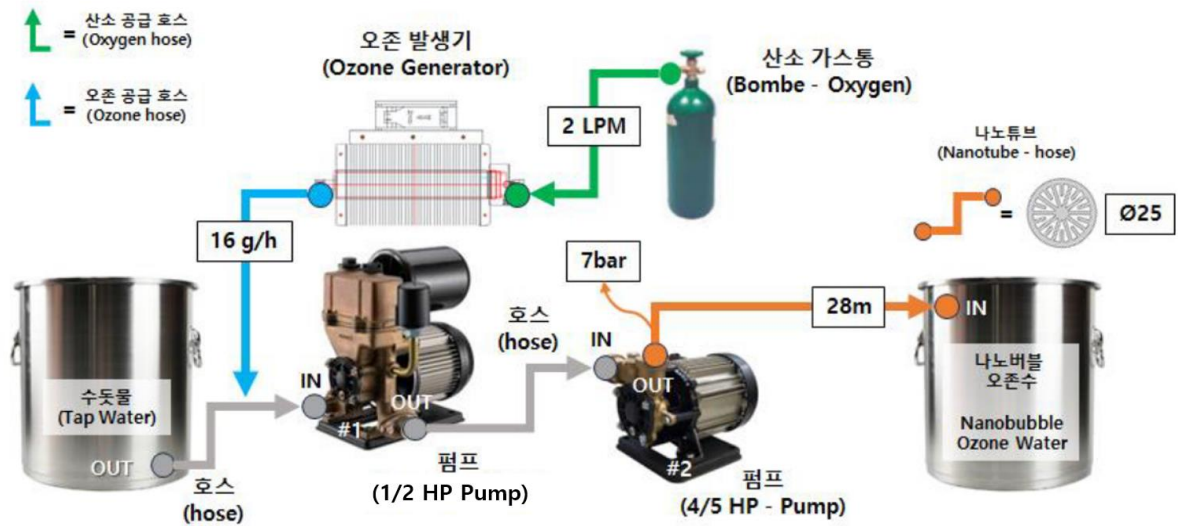

Fig. 2-3 Nanobubble ozone water generation system – for high-concentration ozone water production

To generate high-concentration nanobubble ozone water, the system was constructed using an ozone generator and two pumps. As shown in Fig. 2-3, the sample tank made of SUS 304 was filled with water and the water pump with a rated output of 1/2 HP, a maximum pumping capacity of 2,560 L/h, and a rated power of 220V/60Hz (model: PHH-268A pump (#1), Hanil Electric Co., Ltd.) was prepared to increase flow velocity. The input of the pump was connected to the output of the sample tank using a silicon hose. The output of the #1 pump was then connected to the input of the pump with a rated output of 4/5 HP, a maximum pumping capacity of 3,530 L/h, and a rated power of 220V/60Hz from the same manufacturer (model: MPPH-400 pump (#2)) using a silicon hose.

A nanotube hose with a scale factor of 1.4, an effective friction constant of 39.2, an outer diameter of Ø25, and a length of 28 m was connected to the output of the #2 pump. After confirming a shear pressure of 7.0 bars from the nanotube hose, nanobubble ozone water was produced while the ozone generated through the ozone generator was injected into the input of the #1 pump at a rate of 16 g/h.

The ozone generator used was produced by Ozone Tech (model: OG-16R). To increase the concentration of the ozone generated from the ozone generator, oxygen was supplied into the ozone generator at a rate of 2.0 LPM using the oxygen bombe.

### 3.2 Ozone water produced at a tap water temperature of 14°C and refrigerated

The concentration of the ozone water generated at a tap water temperature of 14°C was measured over time. The same sample was placed in seven 12.5ml glass bottles and immediately stored in a refrigerator. Measurements were then performed through reactions with the reagent (R1) every hour. 6.7 ppm was confirmed at the beginning, and more than 1 ppm was maintained until six hours. After refrigeration, a half-life of three hours was confirmed as the concentration decreased from 2.9 to 1.5 ppm.

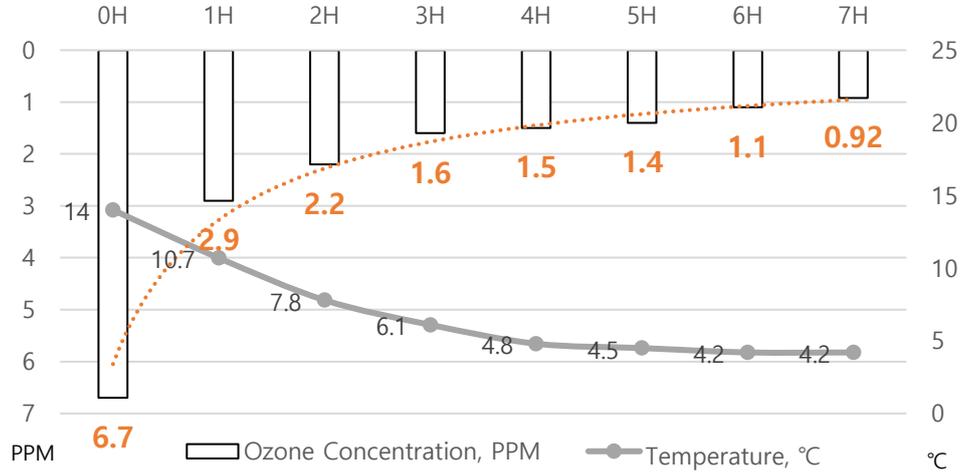

Fig. 2-4 Ozone water concentration change – tap water, room temperature 14°C / refrigeration

### 3.3 Ozone water produced at a tap water temperature of 8°C and refrigerated

The concentration of the ozone water generated at a tap water temperature of 8°C was measured over time. The same sample was placed in seven 12.5ml glass bottles and immediately stored in a refrigerator. Measurements were then performed through reactions with the reagent (R1) every hour. 6.7 ppm was confirmed at the beginning, and 4.9 ppm was maintained in two hours and more than 2 ppm in seven hours. A half-life of approximately three hours and 30 minutes was confirmed as the concentration decreased from 6.3 to 2.8 ppm.

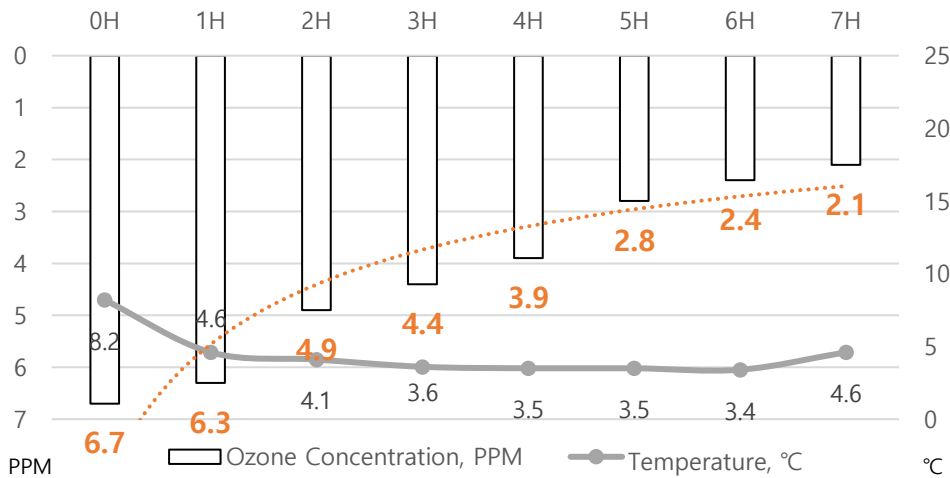

Fig. 2-5 Ozone water concentration change – tap water, room temperature 8°C / refrigeration

### 3.4 Ozone water produced at a tap water temperature of 10°C and rapidly cooled

The concentration of the ozone water generated at a tap water temperature of 10°C was measured over time. The same sample was placed in seven 12.5ml glass bottles and immediately stored in a refrigerator using the rapid cooling function. Measurements were then performed through reactions with the reagent (R1) every hour. Rapid cooling was performed after confirming 10°C_6.5ppm at the beginning. It was found that 3.7°C_5.9 ppm was maintained in an hour and more than 2.5 ppm in seven hours. A half-life of approximately four hours was confirmed as the concentration decreased from 5.9 to 3.1 ppm.

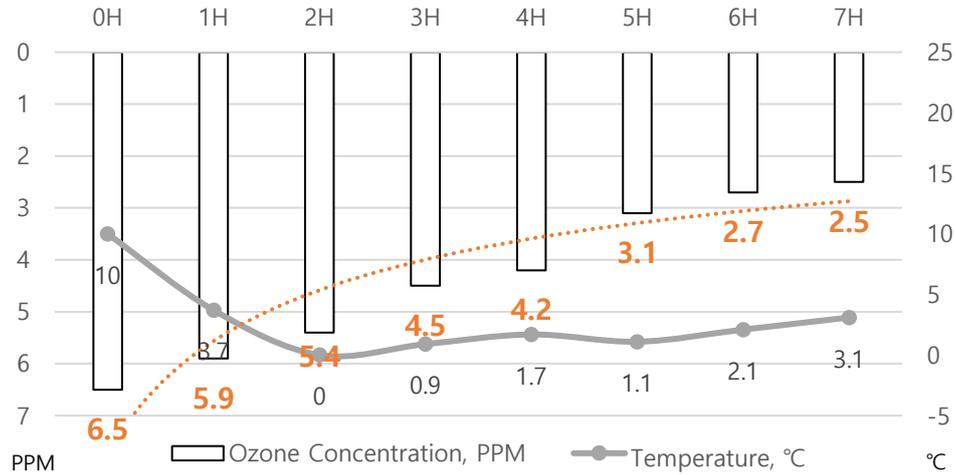

Fig. 2-6 Ozone water concentration change – tap water, room temperature 10°C / rapid cooling

4. **High-concentration nanobubble ozone water production and ozone concentration change - distilled water**

   4.1 **Experimental setup and sample preparation**

High-concentration nanobubble ozone water was generated using distilled water and then half-life was examined and compared. While the laboratory temperature was maintained at 24°C, the experiment was performed using the same system as in Fig. 2-3.

   4.2 **Ozone water generated at a distilled water temperature of 20°C**

The concentration of the ozone water generated at a distilled water temperature of 20°C was measured over time. The same sample was placed in seven 12.5ml glass bottles, and measurements were performed through reactions with the reagent (R1) in new bottles every hour at room temperature. 6.3 ppm was confirmed at the beginning, and it was reduced to a half in approximately two and a half hours. It was found that more than 1 ppm was maintained until seven hours.

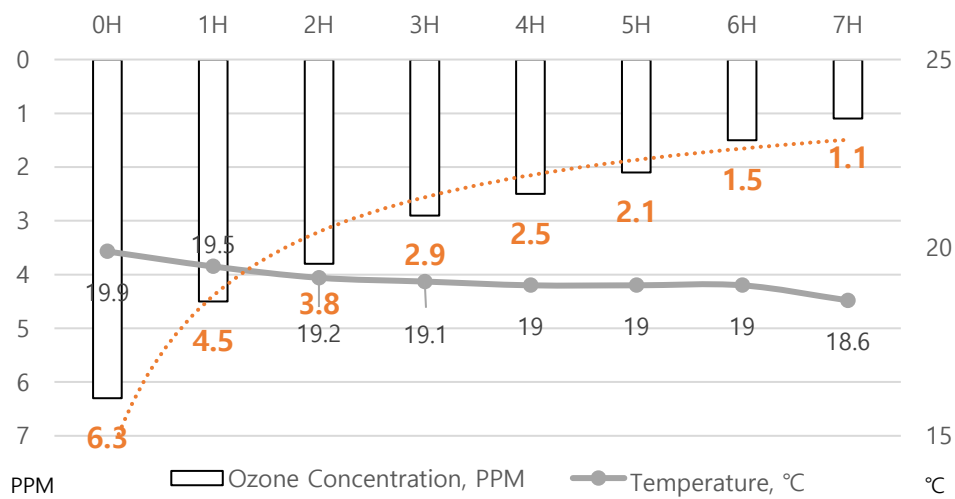

Fig. 2-7 Ozone water concentration change – distilled water, room temperature 20°C / room temperature

### 4.3 Ozone water produced at a distilled water temperature of 10°C and refrigerated

The concentration of the ozone water generated at a distilled water temperature of 10°C was measured over time. The same sample was placed in seven 12.5ml glass bottles and immediately stored in a refrigerator. Measurements were then performed through reactions with the reagent (R1) every hour. After confirming 10°C_6.5 ppm at the beginning, the concentration dropped by approximately 0.4 ppm over time and 3.8 ppm was maintained until seven hours. A half-life of approximately seven and a half hours was observed.

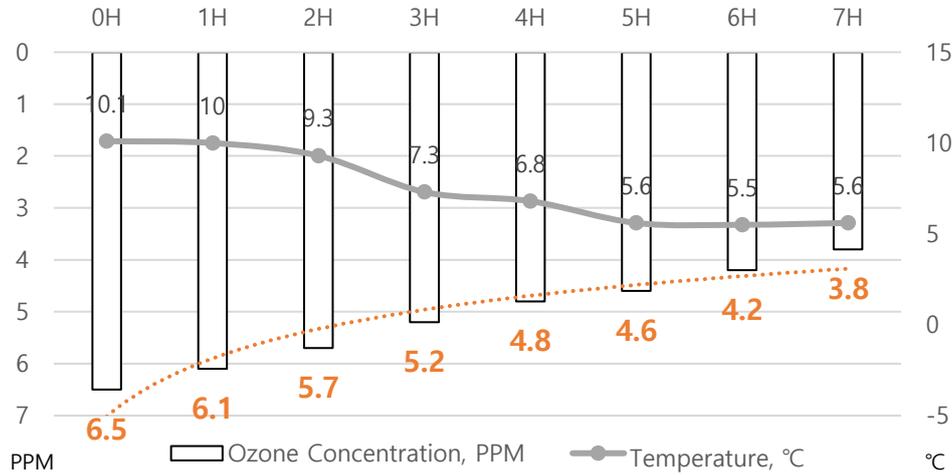

Fig. 2-8 Ozone water concentration change – distilled water, room temperature 10°C / refrigeration

### 4.4 Ozone water produced at a distilled water temperature of 10°C and rapidly cooled

The concentration of the ozone water generated at a distilled water temperature of 10°C was measured over time. The same sample was placed in seven 12.5ml glass bottles and immediately stored in a refrigerator using the rapid cooling function. Measurements were then performed through reactions with the reagent (R1) every hour. Rapid cooling was performed after confirming 10.1°C_6.5ppm at the beginning. It was found that 1.7°C_6.4 ppm was maintained in an hour and more than 4.4 ppm in seven hours.

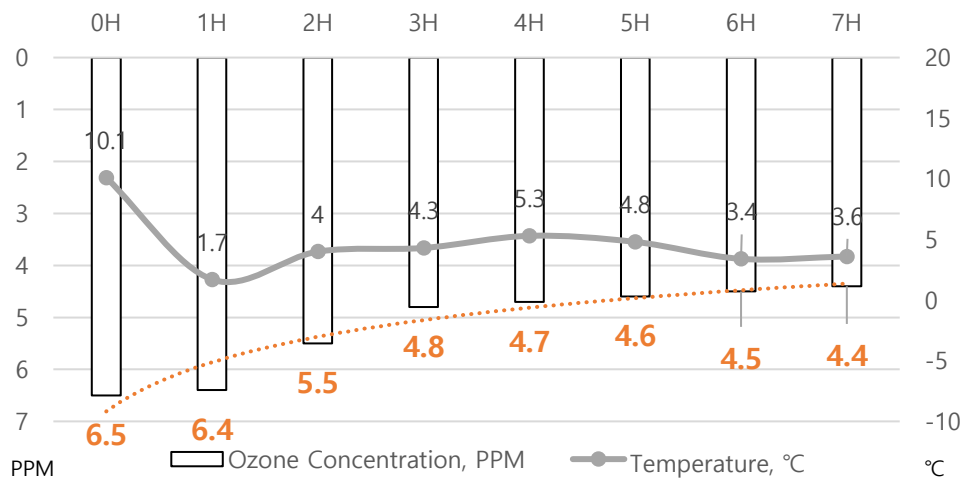

Fig. 2-9 Ozone water concentration change – distilled water, room temperature 10°C / rapid cooling

## 5. Comparison of methods to increase the half-life of nanobubble ozone water

### 5.1 Comparison of experiment results

The measurement results of the ozone water generated at a tap water temperature of 8°C and refrigerated (tap water-refrigeration), the ozone water generated at a distilled water temperature of 20°C and stored at room temperature (distilled water-room temperature), and the ozone water generated at a distilled water temperature of 10°C and rapidly cooled (distilled water rapid cooling-refrigeration) were compared.

It was found that the high concentration could be maintained when the ozone water generated at a low temperature was refrigerated after rapid cooling.

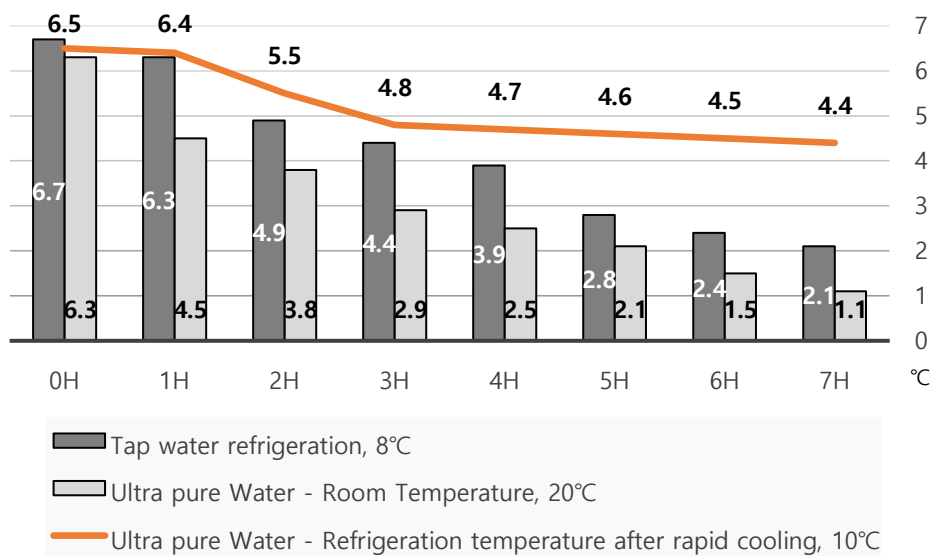

Fig. 2-10 Ozone water concentration comparison - tap water refrigeration, distilled water room temperature, and distilled water rapid cooling

### 5.2 Discussion

The above experiment results show that the sample temperature during the generation of ozone water and the storage temperature after the generation significantly affect the ozone water concentration. When ozone water was generated at a tap water temperature of 8°C and refrigerated at an average temperature of 4.5°C, more than 2 ppm was maintained until seven hours. In the case of distilled water, the standard of ozone water was met by maintaining more than 1 ppm until seven hours even when ozone water was generated at 20°C and stored at room temperature. It was also observed that more than 4 ppm was maintained until seven to eight hours when ozone water was generated at a distilled water temperature of 10°C and immediately refrigerated after rapid cooling.

Based on the above results, it is possible to generate high-concentration ozone water using a nanobubble ozone water generator, and the high-concentration condition can be maintained during distribution by rapid cooling and refrigeration. This can solve the problem that ozone water must be produced and used at the site. Such nanobubble ozone water can be used for the disinfection of medical devices, sterilization of food, disinfection of public facilities, and prevention of seasonal livestock epidemics (e.g. foot-and-mouth disease, African swine fever, and highly pathogenic avian influenza).

### III. Stability of the Ozone Water Sprayed into the Air

#### 1. Test purpose

This test was conducted to examine the change in ozone concentration and the stability of nanobubble ozone water when nanobubble ozone water with a sterilization effect is sprayed into the air.

The environmental standards of ozone are announced by the Ministry of Environment. An eight-hour average of 0.06 ppm and an one-hour average of 0.1 ppm must not be exceeded. In addition, the ozone forecast system classifies the 0.031-0.09ppm range as a normal grade, the 0.091-0.15ppm range as a bad grade, and over 0.151 ppm as a very bad grade.

#### 2. Measurement method

The ozone concentration was measured using the indoor air quality test method of the Ministry of Environment. The ozone concentration in the air was measured for 25 minutes using an ozone monitor (202, 2B Technologies Inc.). For the reliability of the ozone measurement system, its approval by the Ministry of Environment was confirmed through the approval number.

#### 3. Test method and setup

The test was conducted in accordance with the "indoor air quality process standards (December 21, 2018)" announced by the Ministry of Environment under Article 6 (1-3) of the "Act on Testing and Inspection in the Environment Sector". The space for ozone water spray and measurement was classrooms in an elementary school located in Anyang City, Gyeonggi-do, South Korea. The size of each classroom was 166.4 $m^3$ (L 8 m * W 8 m * H 2.6 m).

To generate ozone water using tap water, the same nanobubble ozone water generation system as in Fig. 2-3 was used. Ozone water was generated at an outdoor temperature of 28°C and a tap water temperature of 18°C.

Two ozone water samples to spray were prepared. For the first sample, the ozone water concentration before spray was 5 ppm. 3,600 cc was prepared and sprayed for 90 seconds in a classroom (#1). In this instance, the concentration of ozone water that passed through the injection nozzle was 2.6 ppm. In the case of the second sample, the ozone water concentration before spray was 4.5 ppm. 6,000 cc was prepared and sprayed for 120 seconds in a classroom (#2). In this instance, the concentration of ozone water that passed through the injection nozzle was 2.0 ppm.

At the time of spray, the temperature inside the classrooms was 24°C. All windows and two entrances in each classroom were closed to prevent the inflow of external air. Each classroom was randomly divided into sections by six people to spray ozone water at the same time. The concentration change was monitored in real time using an ozone measuring instrument, and the maximum and average concentrations were measured during the measurement time.

## 4. Test results and discussion

| Ozone concentration (ppm) | Spray concentration (ppm) | Spray amount (cc) | Spray time (sec) | Spray amount per area (cc/$m^3$) | Maximum concentration (ppm) | Average concentration (ppm) |
|---|---|---|---|---|---|---|
| 5.0 | 2.6 | 3,600 | 90 | 21.6 | 0.07 | 0.034 |
| 4.5 | 2.0 | 6,000 | 120 | 36.1 | 0.058 | 0.03 |

Table 3-1. Ozone concentration and maximum/average concentrations (KSD0223200705)

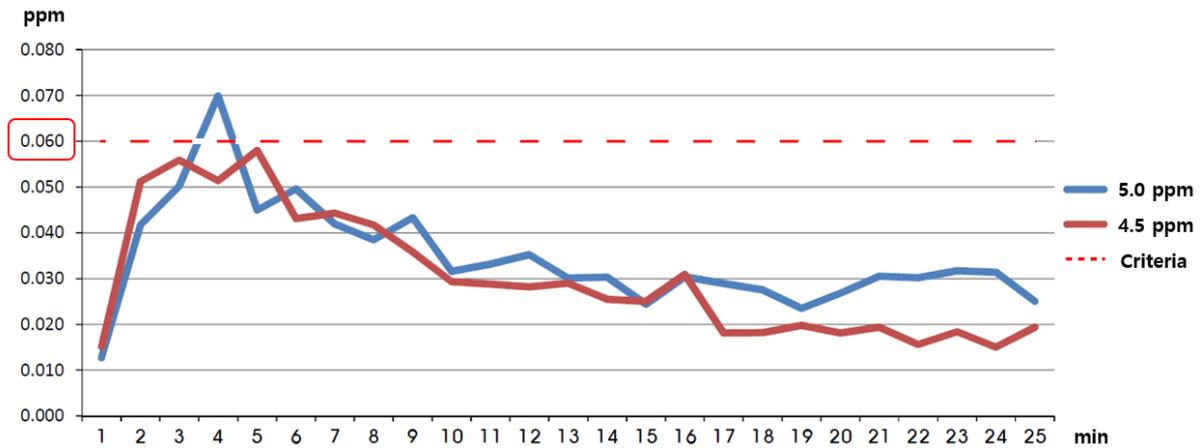

Fig. 3-1 Concentration change after spraying ozone water

In the test results, the average concentration measured was lower than 0.06 ppm, which is the legal limit of ozone, even when high-concentration ozone water was produced and sprayed into the air. In Fig. 3-1, the concentration exceeded the legal limit in three minutes after ozone water spray, but it decreased again within one minute. This is considered safe because it met an eight-hour average of 0.06 ppm and a one-hour average of 0.1 ppm, which are the environmental standards announced by the Ministry of Environment, and it was very short compared to 30 minutes to one hour that affect the human body.

The safety of ozone water was verified by generating ozone water with a concentration that is more than four times higher than 1 ppm, which is the general standard of ozone water, and spraying it into a closed indoor space in large quantities. It is expected that such nanobubble ozone water will be used for spatial disinfection and sterilization by spraying it into the air.

## IV. Virucidal Activity Testing

### 1. Test purpose

To examine whether high-concentration nanobubble ozone water can be used as a disinfectant, influenza A (Influenza A-H1N1/A/PR8) and COVID-19 (Covid-19/SARS-CoV-2) virucidal activity testing was conducted. In the tests, the virucidal activity of nanobubble ozone water was examined in accordance with the U.S. test standard for virus performance evaluation (ASTM E1052-11).

### 2. Experimental setup and sample preparation

To generate ozone water using tap water, the same nanobubble ozone water generation system as in Fig. 2-3 was used. An ozone water concentration of 5.5 ppm was generated at an indoor temperature of 24°C and a tap water temperature of 18°C, and 2.5 ppm was finally confirmed immediately before entry into the laboratory. Afterwards, no additional measurement could be performed due to the laboratory safety issue.

The test on the influenza A virus was conducted in the BSL-2 (No. LML09-180) virus laboratory while the test on the COVID-19 virus was conducted in the BSL-3 (No. KCDC-09-3-01) virus laboratory. These two tests were commissioned to be conducted by 'KR Biotech-Konkuk University'.

### 3. Test method

#### 3.1 Cytotoxicity test method

After elution in accordance with chapter 9 of the common standards on the biological safety of medical devices (Ministry of Food and Drug Safety notice no. 2006-32) or 'ISO10993-12 Sample preparation and reference materials', the test was conducted in accordance with ISO10993-5 or chapter 2 of the common standards on the biological safety of medical devices (Ministry of Food and Drug Safety notice no. 2006-32) without filtering the eluate of the colloid state. In general, cytotoxicity is evaluated using the test method by the direct contact method. This is the test method to find the maximum concentration that does not affect cell survival when cells are cultured during a certain period after continuously diluted samples are processed for them.

The cytotoxicity test was conducted in accordance with the 'Cytotoxicity Test' guidelines (document no. KRBOP-0803-01, crystal violet method) in the 'KR Biotech' laboratory.

#### 3.2 Result observation and judgment method

After diluting the original nanobubble ozone water by 10 and 100 times, the influenza A virus was treated in "MDCK cells" and the COVID-10 virus in the "Vero-E6" cells. For both viruses, only the original ozone water exhibited cytotoxicity compared to the control groups.

In this test, 10-times dilution was performed using the neutralizer (10% FBS) presented in the "disinfectant efficacy test method data package (NIER-GP2018-170)" to neutralize the cytotoxicity of the disinfectant itself.

### 3.3 Disinfectant efficacy test method

Biological tests to evaluate the efficacy of disinfectants need to be conducted in accordance with standard methods, but there are various test methods for disinfectants depending on the target species, method of use, and type. Therefore, it is difficult to uniformly standardize test methods for all disinfectants. In this regard, a test method suitable for the biocide type was selected considering the target organisms, dose, and usage of the biocide.

In this test, the efficacy of the original solution was tested considering the characteristics of ozone water based on ASTM E1052-11.

### 4. Influenza A virus

#### 4.1 Characteristics

Influenza virus diseases are the acute respiratory diseases caused by the influenza virus infection. They involve symptoms, such as sudden chills, high fever, headache, and muscle pain, as the virus infiltrates into the nose, throat, and lungs. Viruses that mainly infect people are influenza A or B.

#### 4.2 Test results

| Sample | Virus $TCID_{50}$ | Processing Time | **Influenza A (H1N1) Virucidal activity** | Virus reduction rate (log) | (%) |
|---|---|---|---|---|---|
| Nanobubble Ozone water | $10^4$ | 5 min | 4/4 | 4 | **99.99%** |
| | | 10 min | 4/4 | | |
| | | 30 min | 4/4 | | |
| | | 60 min | 4/4 | | |
| | $10^5$ | 5 min | 4/4 | 5 | **99.999%** |
| | | 10 min | 4/4 | | |
| | | 30 min | 4/4 | | |
| | | 60 min | 4/4 | | |

Table 4-1. Virucidal activity test results for influenza A (KR-2003-003-FWE01)

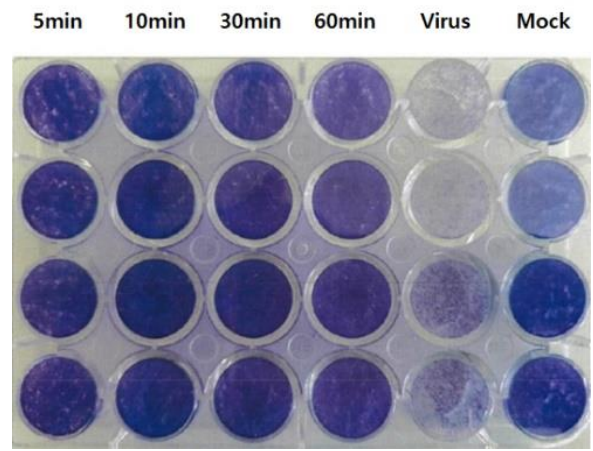

Fig. 4-1 Influenza A (H1N1/A/PR8) $10^4$ $TCID_{50}$ + Nanobubble Ozone Water

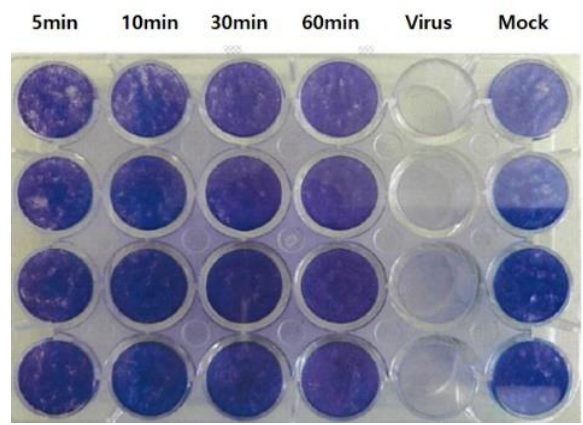

Fig. 4-2 Influenza A (H1N1/A/PR8) $10^5$ $TCID_{50}$ + Nanobubble Ozone Water

Nanobubble ozone water was mixed with the target influenza virus ($10^5$ TCID, $10^4$ TCID)) for reactions over 5, 10, 30 and 60 minutes at room temperature. After diluting it ten times using the neutralizer (10% FBS), it was infected with "MDCK" cells to observe the cytopathic effect caused by the virus. In this instance, a typical saline solution was used as the control group. In the virucidal activity test results of the nanobubble ozone water for influenza A (H1N1/A/PR8), more than 99.99% of the virus was killed by five-minute treatment.

## 5. COVID-19 virus

### 5.1 Characteristics

The COVID-19 virus disease is the respiratory disease caused by the novel coronavirus (SARS-CoV-2). It causes infection when the droplets from an infected person infiltrate through the respiratory system or the mucous membrane of the eyes, nose, and mouth. When infected, the virus causes symptoms, such as fever, cough, and difficulty in breathing, after a certain period of incubation. There are also many infection cases with no symptom.

### 5.2 Test results

| Sample | Virus $TCID_{50}$ | Processing Time | **COVID-19** (SARS-CoV-2) Virucidal activity | Virus reduction rate (log) | (%) |
|---|---|---|---|---|---|
| Nanobubble Ozone water | $10^4$ | 5 min | 4/4 | 4 | **99.99%** |
|  |  | 10 min | 4/4 |  |  |
|  |  | 30 min | 4/4 |  |  |
|  | $10^5$ | 5 min | 4/4 | - | **-%** |
|  |  | 10 min | 4/4 |  |  |
|  |  | 30 min | 4/4 |  |  |

Table 4-2. Virucidal activity test results for COVID-19 (KR-2003-003-FWE01)

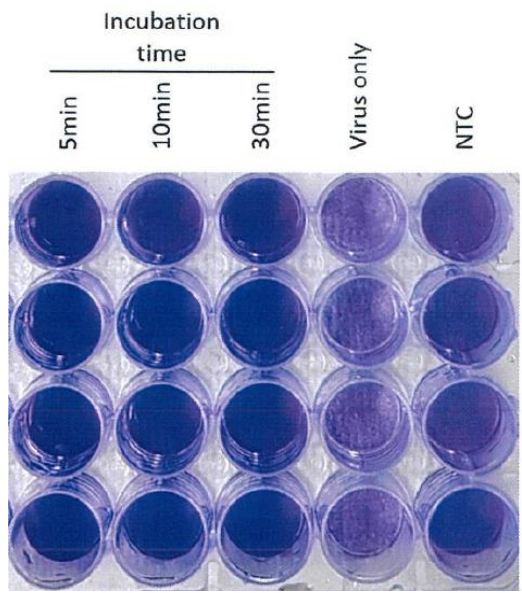

Fig. 4-3 COVID-19 (SARS-CoV-2) $10^4$ + Nanobubble Ozone Water

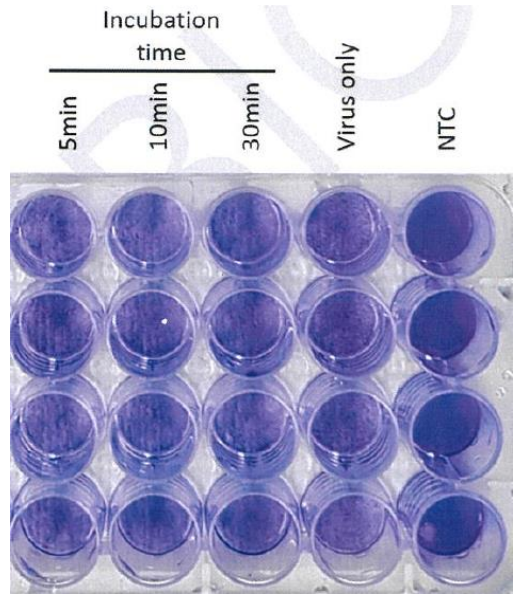

Fig. 4-4 COVID-19 (SARS-CoV-2) $10^5$ + Nanobubble Ozone Water

While the cells infected by the virus (virus only) are not well dyed by "cristal violet" due to their death, cells that are not infected are well dyed as with normal cells (NTC) because they are not infected when the COVID-19 virus is killed by a disinfectant.

Nanobubble ozone water was mixed with the target COVID-19 virus ($10^5$ TCID, $10^4$ TCID)) for reactions over 5, 10, and 30 minutes at room temperature. After diluting it ten times using the neutralizer (10% FBS), it was infected with "Vero-E6" cells to observe the cytopathic effect caused by the virus. In this instance, a typical saline solution was used as the control group. In the virucidal activity test results of the nanobubble ozone water for COVID-19 (SARS-CoV-2), more than 99.99% of the $10^4$ $TCID_{50}$ virus was killed by five-minute treatment.

## V. Results and Discussion

In this study, a system that can generate high-concentration nanobubble ozone water in real time using friction tubes was constructed, and the method of maintaining the concentration over the long term was investigated through experiments. The finding that the stability of atmospheric environmental standards is met when high-concentration ozone water is sprayed into the air for sterilization was verified. In addition, the disinfectant efficacy of high-concentration ozone water against the influenza A virus and COVID-19 virus (SARS-CoV-2) was confirmed.

The results of this study can be summarized as follows.

(1) The use of the nanobubble ozone water generation system makes it possible to produce high-concentration ozone water. It can be distributed in the high-concentration condition when rapidly cooled and refrigerated after production. It is easy to use anywhere and can solve the problem that ozone water must be produced and used at the site.
(2) The average concentration measured was lower than 0.06 ppm, which is the legal limit of ozone, even when high-concentration ozone water was produced and sprayed into the air. The safety of ozone water was verified by generating ozone water with a concentration that is more than four times higher than 1 ppm, which is the general standard of ozone water, and spraying it into a closed indoor space in large quantities.
(3) In the virucidal activity test results of the nanobubble ozone water for influenza A (H1N1/A/PR8) and COVID-19 (SARS-CoV-2), more than 99.99% of the viruses were killed by five-minute treatment.

Various studies have been published so far showing that ozone water is effective in eradicating bacteria.[11] Recently, it was reported that influenza can also be effectively eliminated with ozonated water.[12] Additionally, studies have shown that SARS-CoV-2 can also be removed with ozonated water.[13] However, there was no way to effectively produce ozonated water. The nanobubble ozone water described in this study can be used for healthy and safe sterilization, spatial disinfection, and prevention of epidemics without chemical secondary pollution. It is expected to be further utilized in various industrial sectors.